\documentclass[12pt]{article}
\usepackage{authblk}
\usepackage{graphicx}
\usepackage{epstopdf}
\usepackage{amsmath}
\usepackage{amssymb}
\usepackage{bm}
\usepackage{mathptmx}
\usepackage{xcolor}
\usepackage{tikz}
\usepackage[europeanresistors]{circuitikz}
\usepackage{siunitx}
\usetikzlibrary{shapes.geometric, arrows}
\ifpdf
  \DeclareGraphicsExtensions{.eps,.pdf,.png,.jpg}
\else
  \DeclareGraphicsExtensions{.eps}
\fi
\tikzstyle{startstop} = [rectangle, rounded corners, minimum width=2.5cm, minimum height=2.5cm,text centered, draw=black, fill=blue!30, align=left]
\tikzstyle{decision} = [diamond, minimum width=2cm, minimum height=1cm, text centered, draw=black, fill=green!30]
\tikzstyle{measure} = [thick,->,>=stealth]
\tikzstyle{output} = [coordinate, text centered]

\begin{document}


\markboth{Chialva and Reartes}{On global mechanism of sync and the role of coupling}

\title{On global mechanisms of synchronization in networks of coupled chaotic circuits and the role of the voltage-type coupling}

\author[$1,2$]{Ulises Chialva}
\author[$1$]{Walter Reartes}

\affil[$1$]{Departamento de Matem\'atica, Universidad Nacional del Sur, Av. Alem 1253\\
8000 Bah\'ia Blanca, Buenos Aires, Argentina\\
uchialva@gmail.com}

\affil[$2$]{CONICET}

\maketitle

\begin{abstract}
A model for synchronization of coupled Nakano's chaotic circuits is studied. The Nakano circuit consists of a simple RLC circuit with a switch voltage-depending reset rule which generates a discontinuous dynamics. Thus, the model that we study is a network of identical spiking oscillators with integrate-and-fire dynamics. The coupling between oscillators is linear, but the network is subject to a common regime of reset depending on the global state of the oscillator population. This constitutes the simplest way of build pulse-coupled networks with arbitrary topology for this type of oscillators, and it allows the emergence of synchronous states and different reset regimes.

The main result is that under certain hypothesis over the weight matrix (that represents the network topology) the different reset regimes match and the formalism of the master stability function can be generalized in order to study the stability of the synchronous state and the discontinuous dynamic of the network. Also, the low dimensionality of the Nakano's circuit allows to implement the saltation-matrix method and numerical simulations can be performed in order to analyze the role of the coupling mode in the synchronization regime of the network and the influence of the voltage-type variables.
\end{abstract}


\section{Introduction}
The integrate-and-fire models (IFMs) are known to be one of the simplest neuronal models, which in turn replicate a large number of present behaviors in biological neurons, such as spikes, bursting, mode-locked states, etc. \cite{izhikevich08,coombes99}. At the same time, using the
IFMs, a specific type of networks called pulse-coupled networks can be constructed. The analysis of the synchrony of these networks is important not only as basic nonlinear problems but also as an approach to many biological systems as pacemaker cells of the heart \cite{peskin1975mathematical}, flashing of fireflies populations \cite{mirollo1990synchronization}, segregation of insulin in the pancreas \cite{sherman1989collective}, etc.

On the other hand, consideration of electric circuit versions of integrate-and-fire systems is important in view of obtaining more realistic and plausible systems of experimental implementations. Also the synchronous phenomenon in pulse-coupled network of these oscillators has been studied due to its relation with models of voltage-coupled cells \cite{mancilla2007synchronization, chartrand2019sync} and some engineering applications, for example chaos-based communications systems \cite{argyris05} and artificial neural networks \cite{hoppensteadt00}.

This paper concerns the emergence of synchrony in a specific pulse-coupled network. Our work was inspired by the Nakano's chaotic circuit \cite{nakano02}, which consists of a simple $\mathcal{RLC}$ circuit with a self-regulating switch. Despite its simplicity, the circuit presents complex behaviors as bursting states and chaotic oscillations. In turn, the system that describes the temporal evolution of the circuit is one of the simplest two-dimensional chaotic discontinuous systems. In the past, the Nakano's oscillator was studied as an isolated node or coupled in the limited cases of a \emph{master-slave} configuration of two nodes or in ring configurations. In this work, we generalize that type of coupling by considering networks that have arbitrary number of nodes and any desired topology. This network model guarantees the existence of a synchrony solution and allows to raise the problem of its stability.

In recent years, the application of the master stability function (MSF) \cite{pecora98,pecora00} to the study of synchronization in discontinuous oscillator networks has begun to be of interest, and its adaptation has been carried out for a wide family of neural integrate-and-fire models \cite{coombes16,coombes12,ladenbauer2013}. At the same time, problems such as the appearance of different reset regimes in the nodes and their influence on the variational equation \cite{nicks18, csayli2019synchrony} or the prediction of synchronization clusters have required expanding the used formalism.

The main goal of this work is the construction of a network model that presents the necessary conditions to expand the classic MSF in this new context. The features of the Nakano's model allow the construction of a network whose structure guarantees conditions to control the different reset regimes of its variational equation. In this way it is possible to consider the formalism of the MSF for networks that have a matrix of weights satisfying a specific condition (that we have called ``doubly balanced"). In turn, the low dimensionality of the Nakano's model allows the implementation of the saltation-matrix method \cite{muller95,dibernardo08,kunze2000} to evaluate the exponents of Lyapunov required by the MSF. Finally, the numerical stability of these evaluations makes it possible to compare the influence of the coupling regime (and in particular those that involve the voltage-type variables) through its MSF corresponding to the symmetric networks.

In Section \ref{nakano_model} we show the Nakano circuit and the impact system that describes its temporal evolution. We also introduce the \emph{dummy} coupling and its generalization for arbitrary topologies. In Section  \ref{red_hibrida_seccion} we introduce the hybrid networks of an arbitrary number of nodes subject to a common restart, and show the existence of a synchrony solution and show how the approach of the variational equation gives rise to different reset regimes. Also, we define a condition that guarantees the control of such regimes. In Section \ref{extension_formalismo} we use the mentioned hypotheses and expand the MSF for this new type of network. In \ref{evaluacion_de_MSF}, through the \emph{saltation matrix} method, we give a method to evaluate the exponents of Lyapunov and the MSF of the Nakano oscillator. Finally, in Section \ref{seccion_acoples}, through the numerical simulations implemented with the methods of the previous sections, we investigate the influence of the different types of coupling, and exhibiting the stabilizing effect that the couplings through the variable voltage-type possess.

\section{The Nakano-Saito model}
\label{nakano_model}
\begin{figure}
\begin{center}
\begin{circuitikz}{european} \draw
(6,0) --(0,0)to[R, l_= -$\mathcal{R}$](0,3) to[inductor, l_ = $\mathcal{L}$, i = $i$] (3,3)-- (6,3)--(6,2) to[cspst,l=$S$] (6,1);
\draw[dashed](-0.5,-0.5)rectangle(2.75,3.75); 
\draw (6,0) to[battery1,l_=  $\mathcal{E}$] (6,1);
\draw (6,0)--(4,0) to[capacitor,l_ = $\mathcal{C}$] (4,3) -- (4,3.675);
\draw (5,4) node[op amp, scale = 0.65](opamp){}
 (opamp.+)-- (4,3.676)
 (opamp.-)--(4,4.314)node[anchor=east]{$V_T$}
 (opamp.out)-- (6.5,4);
\node[draw,minimum width=0.5cm,minimum height=0.75cm,anchor=south west] at (6.5,3.65){MM};
\draw (6,4.75) node[anchor=east]{COMP}(6,4.75);
\draw (3.5,2.5) node[anchor=east]{+}(3.5,2.5);
\draw (3.5,1.5) node[anchor=east]{$v$}(3.5,1.5);
\draw (3.5,0.5) node[anchor=east]{-}(3.5,0.5);
\draw (7.4,4)--(8,4)--(8,1.5)--(6.2,1.5);
\label{circuito-esquema}
\end{circuitikz}
\caption{Model of Nakano-Saito circuit given by the system \eqref{circuito}. In the figure, $-\mathcal{R}$, COMP and MM denote a linear negative resistor, a comparator and a monostable multivibrator, respectively. If the capacitor voltage reaches the threshold $V_T$, the COMP triggers the MM to output a signal that closes the switch $S$ and $v$ is reset to base voltage $\mathcal{E}$.}
\end{center}
\end{figure}
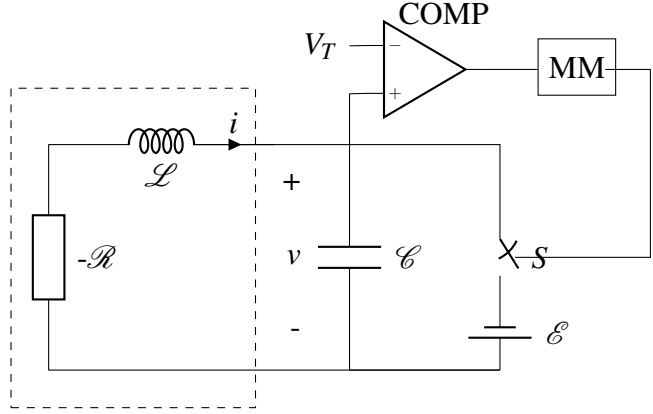

We consider the following integrate-and-fire model introduced by Nakano and Saito in \cite{nakano02}, it consists of a $\mathcal{RLC}$ circuit with a switched regime given by a condition on the voltage variable:
\begin{equation}
\left\{
\begin{aligned}
& \mathcal{C}\frac{dv}{dt}=i\\
& \mathcal{L}\frac{di}{dt}=-v+\mathcal{R}i\\
& v(t^-)=\mathcal{V}_T\rightarrow (v(t^+),i(t^+))=(\mathcal{E},i(t^-)).\\
\end{aligned}
\right.
\label{circuito}
\end{equation}
$v,i$ are the voltage and current variables, $\mathcal{R}$, $\mathcal{L}$, $\mathcal{C}$ are the constants corresponding to the resistance, the inductance and the capacitor respectively. $\mathcal{V}_T$ and $\mathcal{E}$ are the threshold and base voltage for the switch condition. In Figure \ref{circuito-esquema} the circuit model is showed.

Suppose that \eqref{circuito} has unstable complex characteristic root $\rho\omega+\textrm{i}\omega$
\begin{equation}
\omega^2=\frac{1}{\mathcal{L}\mathcal{C}}-\left(\frac{\mathcal{R}}{2\mathcal{L}}\right)^2>0,\quad\rho=\frac{\mathcal{R}}{2\omega\mathcal{L}}>0.
\end{equation}
Now, we use the following dimensionless variables and parameters:
\begin{equation}
\begin{aligned}
\tau &= \omega t\\
v_R &= \frac{\mathcal{E}}{\mathcal{V}_T}\\
x_1 &=\frac{v}{\mathcal{V}_T}\\
x_2 &=-\frac{\rho}{\mathcal{V}_T}v+\frac{1}{\omega\mathcal{C}\mathcal{V}_T}i.\\ 
\end{aligned}
\end{equation}
For brevity we write $\bm x^\pm=\lim_{\epsilon\to 0^\pm}\bm x(t+\epsilon)$.
The derivatives are with respect to $\tau$. So, the system \eqref{circuito} is transformed into
\begin{equation}
\left\{
\begin{aligned}
&\dot{\bm x} = A\bm x\\
& x_1^-=1\rightarrow\bm x^+=R\bm x^-+I,
\end{aligned}
\right.
\label{sistema_nakano}
\end{equation}
where
\begin{equation*}
\bm x = \begin{pmatrix}x_1\\x_2\end{pmatrix},\quad
A = \begin{pmatrix}
\rho & 1\\
-1 & \rho
\end{pmatrix},\quad
R = \begin{pmatrix}
0 & 0\\
0 & 1
\end{pmatrix},\quad
I = \begin{pmatrix}
v_R\\
\rho (1-v_R)
\end{pmatrix}.
\end{equation*}
\begin{figure}
\includegraphics[scale = 0.4]{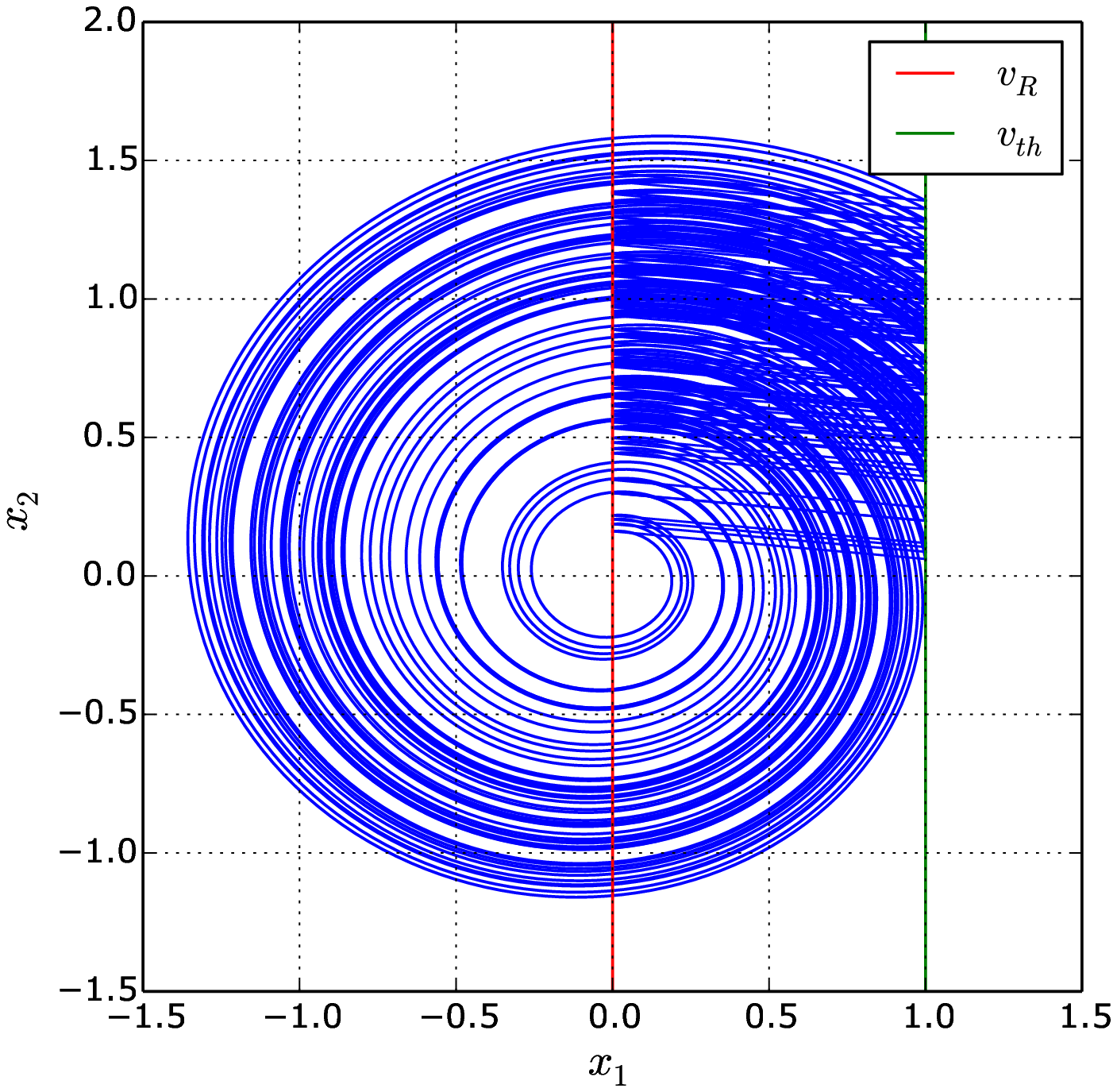}
\includegraphics[scale = 0.3]{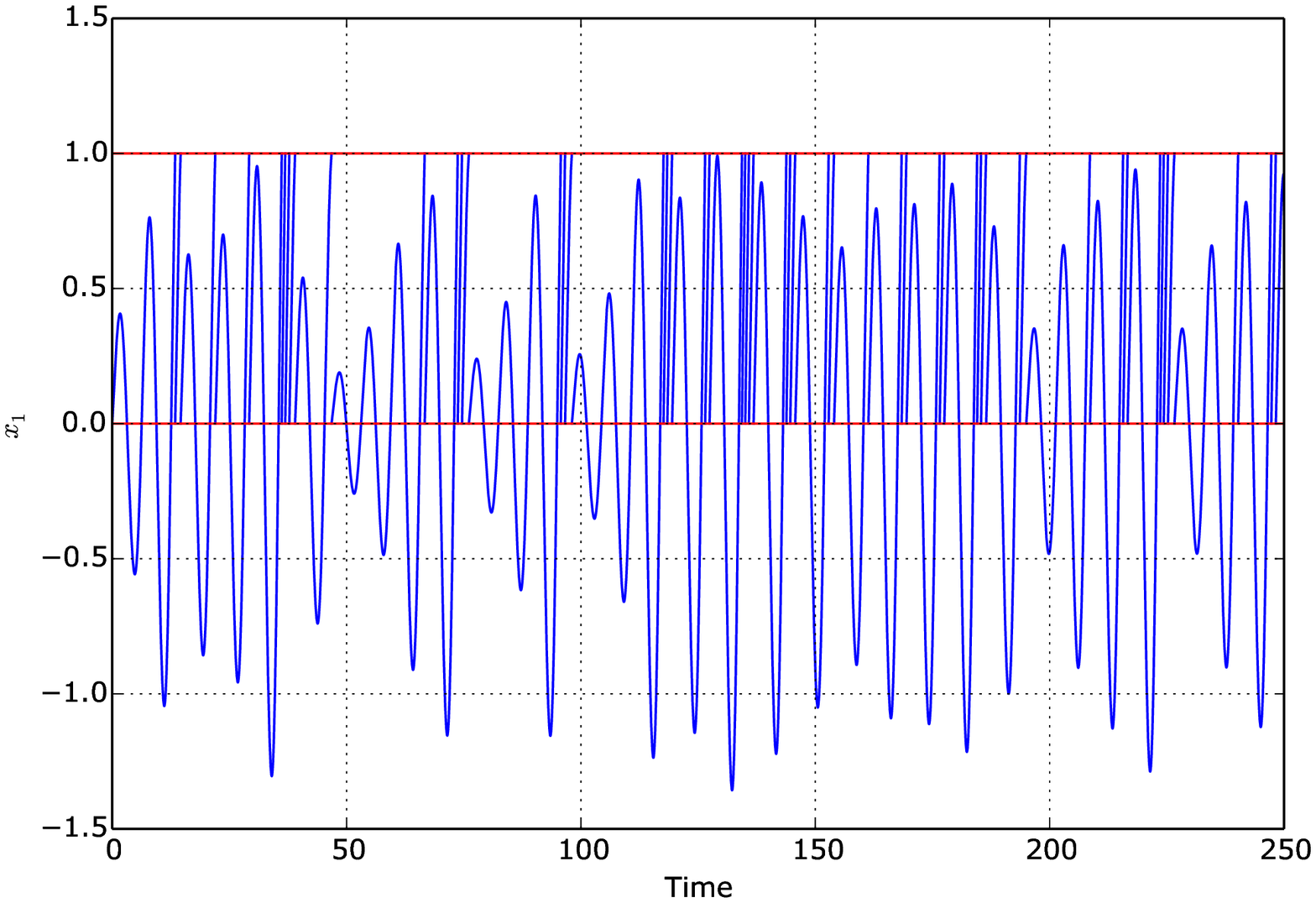}
\caption{Nakano-Saito chaotic oscillator (parameters: $\rho = 0.1$, $v_R=0$).}
\label{nakano-saito-oscilador}
\end{figure}
System \eqref{sistema_nakano} has an unstable equilibrium of focus type in 0, and exhibits bursting states. In particular, the divergent vibration and the firing switch acts as stretching and folding mechanisms, respectively, which are fundamental for chaos generation. In \cite{nakano02} the chaotic behavior of system \eqref{sistema_nakano} was proved, and the synchrony of two coupled systems in a \emph{master-slave} configuration was investigated too, see Figure \ref{nakano-saito-oscilador}. In addition, as is usual in integrate-and-fire systems, the variable that is selected for the reset condition is called the \emph{voltage-type} variable \cite{chartrand2019sync}. In this case it is variable $\bm x_1$, which is proportional to the variable that measures the voltage in system \eqref{circuito}.

We are interested in the collective behavior that emerges from a particular coupling way, called \emph{dummy slave}. Consider a master node $\bm x=(x_1,x_2)$ given by system \eqref{sistema_nakano}, and a slave node $\bm y =(y_1,y_2)$ coupled as follow:
\begin{equation}
\left\{
\begin{aligned}
&\dot{\bm y} = A\bm y+E\bm x\\
& x_1^-=1\rightarrow\bm y^+=R\bm y^-+I,
\end{aligned}
\right.
\label{sistema_dummy}
\end{equation}
where $E\in\mathbb{R}^{2\times 2}$. The previous system is the simplest way to couple these oscillators and is used as a first approach to study other more intricate couplings (see again \cite{nakano02}). So, as approach to general case, we consider networks of arbitrary number of nodes mutually coupled in that way. This way of coupling, we will see, allows a simpler reset regime.

Our generalization of the \emph{dummy-slave} coupling allows us to obtain a network with a single reset regime. In turn, this allows us to study how certain inputs that measure the global state of nodes influence the stability of synchrony solution.

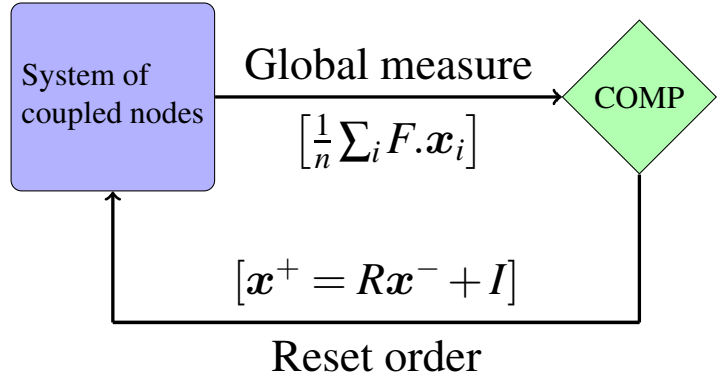
\begin{figure}
\begin{center}
\begin{tikzpicture}[node distance=3cm]
\node (start) [startstop] {System of \\ coupled nodes};
\node (COMP) [decision, right of=start, xshift=4cm] {COMP};
\draw [->,line width=1.25pt] (start) --node[above = 2pt] {\Large Global measure} (COMP);
\draw [->] (start) --node[below = 2pt] {\Large $\left[\frac{1}{n}\sum_iF.\bm x_i\right]$} (COMP);
\node [output, below of=COMP] (vuelta1) {};
\node [output,below of = start] (vuelta2) {};
\draw [-,line width=1.25pt] (COMP) -- node [name=y] {}(vuelta1);
\draw [-,line width=1.25pt] (vuelta1) -- node [above = 2pt] {\Large $\left[\bm x^+=R\bm x^-+I\right]$}(vuelta2);
\draw [-,line width=1.25pt] (vuelta1) -- node [below = 2pt] {\Large Reset order}(vuelta2);
\draw [->,line width=1.25pt] (vuelta2) -- node [name=y] {}(start);
\end{tikzpicture}
\caption{Symbolized diagram of the network model with a global feedback. The system of coupled nodes send a measure of the global activity and it is analyzed by the COMP element. If certain condition is fulfilled (to reach certain threshold for example), so COMP sends a reset order to the nodes.}
\end{center}
\label{esquema_red}
\end{figure}

The key to our model is to take a measure of the global behavior of the network (in our case it is the average of the \emph{voltage-type} variables). The value of this measure is received by an external comparator that evaluates its magnitude. When a certain value is reached, the comparator issues the network reset order. For simplicity, we maintained an affine linear function as reset rule. See Figure \ref{esquema_red}. As we will see, this global feedback allows the appearance of different synchrony regimes.

\section{Hybrid Recurrent Networks}
\label{red_hibrida_seccion}
We consider the next \emph{hybrid network} (we use such name because the system is hybrid in the nomenclature of \cite{dibernardo08}) of $n$ linear oscillators $\bm x_i\in\mathbb{R}^2$, with $A\in\mathbb{R}^{2\times 2}$, $L=(l_{ij})\in\mathbb{R}^{n\times n}$ a laplacian matrix (that is, the sum over each row is zero, i.e. the network is balanced) and the  matrix $E\in\mathbb{R}^{2\times 2}$, which captures the way in which information among nodes is being exchanged by identifying what states a node transmits to its neighbors:
\begin{equation}
\left\{
\begin{aligned}
& \dot{\bm x}_i = A \bm x_i+\sum_{j=1}^n l_{ij} E\bm x_j\\
& \frac{1}{n}\sum_{k=1}^n F.\bm x_k^-=1\rightarrow \bm x_j^+=R\bm x_j^-+I,
\end{aligned}
\right.
\label{red_hibrida}
\end{equation}
also $F\in\mathbb{R}^2$, $R\in\mathbb{R}^{2\times 2}$ is the \emph{reset} matrix and the vector $I\in\mathbb{R}^2$ is called \emph{input}.

Suppose now that there is a synchronous state (or synchrony solution) $\bm s(t)=\bm x_1(t)=...=\bm x_n(t)$. Because the sum of each row of the $L$ matrix is zero, such solution is given by the system
\begin{equation}
\left\{
\begin{aligned}
& \dot{\bm s}=A\bm s\\
&F.\bm s^-=1\rightarrow \bm s^+=R \bm s^-+I.
\end{aligned}
\right.
\label{solucion_sincronia}
\end{equation}

Note that if we replace the values of matrices $A$, $E$, $I$ with those given in the previous section and take $F=(1,0)$, then the synchrony solution is solution of Nakano's system.

Now consider the \emph{average variable} $\bm u =:\frac{1}{n}\sum_i \bm x_i$, and write the variational equation around $\bm s(t)$: consider the variation $\delta\bm x_i=\bm x_i-\bm s$. Using that
$$\sum_{j=1}^nl_{ij}(E\bm x_i-E\bm s)=\sum_{j=1}^nl_{ij}E\delta\bm x_j,$$ and $\bm s^+=R\bm s^-+I$, we have the next variational system ($\mathbb{I}_2$ is the identity matrix of dimension 2):
\begin{equation}
\left\{
\begin{aligned}
&\delta\dot{\bm x}_i=A\delta\bm x_i+\sum_{j=1}^nl_{ij}(E\bm x_i-E\bm s)=A\delta\bm x_i+\sum_{j=1}^nl_{ij}E\delta\bm x_j\\
&F.\bm s^-=1\rightarrow\delta\bm x_i^+=\delta\bm x_i^-+(\mathbb{I}_2-R)\bm s^--I\\
&F.\bm u^-=1\rightarrow\delta\bm x_i^+ = R\delta\bm x_i^--(\mathbb{I}_2-R)\bm s^-+I.\\
\end{aligned}
\right.
\label{ecuacion_variacional_grande}
\end{equation}
We observe that there are two conditions (or regimes) for resetting the systems, and each condition has its own reset rule.

Similar scenarios can be found in \cite{nicks18}. There, the problem of different regimes is approached by considering sums of mass-point functions for the instants of reset and re-writing the system to introduce them. Here, we consider an alternative approach.

We introduce an additional condition on the network topology: we will ask that the sum along each column of the weights matrix $L$ is also zero (we say that the matrix is \emph{doubly balanced} in this case). This condition contemplates the traditional case of non-directed graphs (in this case, the matrix $L$ is symmetric).

Assuming the network is doubly balanced, we have the following: 
\begin{equation}
\begin{split}
\dot{\bm u}=\frac{1}{n}\sum_{i=1}^n\dot{\bm x}_i=& A\bm u +\frac{1}{n}\sum_{i=1}^n\sum_{j=1}^n l_{ij}E\bm x_j\\
&= A\bm u+\frac{1}{n}\sum_{j=1}^n\left(\sum_{i=1}^n l_{ij}\right)E\bm x_j\\
& = A\bm u.
\end{split}
\end{equation}
In addition the reset rule is transformed into the following expression:
\begin{equation}
\frac{1}{n}\sum_{j=1}^nF.\bm x_j^-=F.\bm u^- = 1\rightarrow \bm u^+=R\bm u^-+I.
\end{equation}
Finally, we obtain that the average solution obeys the system
\begin{equation}
\left\{
\begin{aligned}
&\dot{\bm u}=A\bm u\\
&F.\bm u^- = 1\rightarrow \bm u^+=R\bm u^-+I.
\end{aligned}
\right.
\label{solucion_promedio}
\end{equation}
But observing that the systems  \eqref{solucion_sincronia} and \eqref{solucion_promedio} are the same, we get that \emph{for a doubly balanced network its synchrony solution matches with the averages of nodes}. So, we conclude that the structure of the network (represented through the matrix $L$) produces the coincidence of the reset regimes.
\section{Extension of MSF}
\label{extension_formalismo}
If $\bm u = \bm s$, then the variables $\bm x_i$ and $\bm s$ reset at the same time by the same rule. so we will have that system \eqref{ecuacion_variacional_grande} becomes in
\begin{equation}
\left\{
\begin{aligned}
&\delta\dot{\bm x}_i=A\delta\bm x_i+\sum_{j=1}^nl_{ij}(E\bm x_i-E\bm s)=A\delta\bm x_i+\sum_{j=1}^nl_{ij}E\delta\bm x_j\\
&F.\bm s^-=1\rightarrow\delta\bm x_i^+=R\delta\bm x_i^-.
\end{aligned}
\right.
\label{ecuacion_variacional}
\end{equation}

Note that the previous system is an hybrid system whose reset condition is independent of the state of the nodes and only depends on the frequency of $\bm s(t)$. That is, the evolution of variation and resetting are ``decoupled".

Now we write everything with the usual Kronecker product. Let be $\delta\bm x=(\delta\bm x_i,...,\delta\bm x_n)$. We obtain
\begin{equation}
\left\{
\begin{aligned}
&\delta\dot{\bm x}=(\mathbb{I}_n\otimes A)\delta\bm x+(L\otimes E)\delta\bm x\\
&F.\bm s^-=1\rightarrow\delta\bm x^+=(\mathbb{I}_n\otimes R)\delta\bm x^-.
\end{aligned}
\right.
\end{equation}

Let $P\in\mathbb{R}^{n\times n}$ be such that $L=P\Lambda P^{-1}$, where $\Lambda$ is diagonal, and make the change of variables $(P\otimes\mathbb{I}_2)\bm\xi =\delta\bm x$.

So, we have
\begin{equation}
(P\otimes\mathbb{I}_2)\dot{\bm\xi}=(\mathbb{I}_n\otimes A)(P\otimes\mathbb{I}_2)\bm\xi+(P\Lambda P^{-1}\otimes E)(P\otimes\mathbb{I}_2)\bm\xi.
\end{equation}
Using properties of the Kronecker product we finally get
\begin{equation}
\dot{\bm\xi}=(\mathbb{I}_n\otimes A)\bm\xi+(\Lambda\otimes E)\bm\xi.
\end{equation}
For the reset conditions we have
\begin{equation}
\begin{aligned}
F.\bm s^-=1\rightarrow\bm\xi^+&=(P\otimes\mathbb{I}_2)^{-1}(\mathbb{I}_n\otimes R)(P\otimes\mathbb{I}_2)\bm\xi^-\\
&=(P^{-1}\otimes\mathbb{I}_2)(P\otimes R)\bm\xi^-\\
&=(\mathbb{I}_n\otimes R)\bm\xi^-.
\end{aligned}
\end{equation}

Finally, knowing that the matrix $\Lambda$ is diagonal, we can decouple the different normal modes in the usual way. Let be $\lambda_i = \alpha+\textrm{i}\beta$ and $\bm\xi_i=\bm\eta$, we have that each block obeys the equation:
\begin{equation}
\left\{
\begin{aligned}
&\dot{\bm\eta}=A\bm\eta+(\alpha+\textrm{i}\beta)E\bm\eta\\
&F.\bm s^-=1\rightarrow \bm\eta^+=R\bm\eta^-,
\end{aligned}
\right.
\label{sistema_desacople_MSF}
\end{equation}
where the orbit $\bm s$ is given by the system \eqref{solucion_sincronia}.

In this way, we can use the system \eqref{sistema_desacople_MSF} to define the MSF in the usual way as the maximum Lyapunov exponent for given $\lambda=\alpha+\textrm{i}\beta$, and thus we can analyze the stability of the synchronous state.
\section{Evaluating the MSF}
\label{evaluacion_de_MSF}
For evaluating the MSF, we join the previous system \eqref{sistema_desacople_MSF} with \eqref{solucion_sincronia}:
\begin{equation}
\left\{
\begin{aligned}
&\dot{\bm s}=A\bm s\\
&\dot{\bm\eta}=A\bm\eta+(\alpha+\textrm{i}\beta)E\bm\eta\\
&F\bm s^-=1\rightarrow
\begin{aligned}
\bm s^+&=R\bm s^-+I\\
\bm \eta^+&=R\bm\eta^-
\end{aligned}
\end{aligned}
\right.
\label{sistema_MSF}
\end{equation}

We take system \eqref{sistema_MSF} and use a more compact notation, where $\lambda = \alpha+\textrm{i}\beta$:
\begin{equation*}
\bar{A} = 
\begin{pmatrix}
A & 0\\
0 & A+\lambda E
\end{pmatrix},\quad
\bar{R}=
\begin{pmatrix}
R & 0\\
0 & R
\end{pmatrix},\quad
\bar{I}=
\begin{pmatrix}
I\\
0
\end{pmatrix},\quad
\bar{F}=
\begin{pmatrix}
F\\0
\end{pmatrix},
\end{equation*}
where 0 represents null-matrices with adequate size. 

Now we define the variable $\bar{\bm\eta} = (\bm s,\bm\eta)^T$. In this way, we write the system \eqref{sistema_MSF} in a more compact way:
\begin{equation}
\left\{
\begin{aligned}
&\dot{\bar{\bm\eta}}=\bar{A}\bar{\bm\eta}\\
&\bar{F}.\bar{\bm\eta}^-=1\rightarrow\bm\eta^+=\bar{R}\bm\eta^-+\bar{I}.
\end{aligned}
\right.
\label{sistema_compacto}
\end{equation}

Let $t_k,k\in\mathbb{N}$ be the instants where there is a reset of variable $\bar{\bm\eta}$, and let $\mathcal{Q}(t)$ be the \emph{saltation matrix} of the system \eqref{sistema_compacto} (see Chapter 2 of \cite{dibernardo08}). Write $G(t)=\exp{\bar{A}t}$, so we have the next approximation to the variation $\delta\bar{\bm\eta}$:
\begin{equation}
\delta\bar{\bm\eta}(t_n)=\mathcal{Q}(t_n)G(t_{n}-t_{n-1})...\mathcal{Q}(t_1)G(t_1-t_0)\delta\bar{\bm\eta}(t_0).
\label{variacion_matriz}
\end{equation}

But since we want to evaluate the maximal Lyapunov coefficient of \eqref{sistema_desacople_MSF}, we consider only initial variations of the form
\begin{equation}
\delta\bar{\bm\eta}(0)=
\begin{pmatrix}
0\\
...\\
0\\
\delta\bm\eta_1(0)\\
...\\
\delta\bm\eta_m(0)\\
\end{pmatrix}
\label{variacion_inicial}
\end{equation}

and use the formula \eqref{variacion_matriz} to approximate the limit
\begin{equation}
\Lambda_{\max}=\lim_{k\to\infty}\frac{1}{t_k}\log\left(\frac{\Vert\delta\bar{\bm\eta}(t_k)\Vert}{\Vert\delta\bar{\bm\eta}(0)\Vert}\right).
\label{formula_lyapunov_vert}
\end{equation}

\subsection{The saltation matrix}
Let's write $[A,B]=AB-BA$, where $A,B$ are square matrices with the same size. Also, write $\bar{\bm\eta}_*=(\bm s_*,\bm\eta_*)$ the cut points in which the variable $\bar{\bm\eta}$ is reset. Then, according to the given formula for the \emph{saltation matrix} $\mathcal{Q}$ of \eqref{sistema_compacto} (see the Section 2.5 of \cite{dibernardo08}), we have:

\begin{equation}
\begin{aligned}
\mathcal{Q}(\bar{\bm\eta}_*)&=\bar{R}+\frac{\left(\bar{A}\bar{R}\bar{\bm\eta}_*+\bar{A}\bar{I}-\bar{R}\bar{A}\bar{\bm\eta}_*\right)\bar{F}^\top}{\bar{F}^\top\bar{A}\bar{\bm\eta}_*}\\
&= 
\begin{pmatrix}
R & 0\\
0 & R\\
\end{pmatrix}
+
\frac{1}{F^\top A\bm s_*}
\begin{pmatrix}
[A,R]\bm s_*F^\top+IF^\top & 0\\
[A+\lambda E,R]\bm\eta_*F^\top & 0\\
\end{pmatrix}.
\end{aligned}
\label{saltation_matrix}
\end{equation}

We only conserve the fourth block of the matrix \eqref{saltation_matrix} because we use the formula \eqref{variacion_matriz} with an initial variation \eqref{variacion_inicial}:
\begin{equation}
\mathcal{Q}'=R,
\end{equation}
Let be $G'(t)=\exp{(A+\lambda E)t}$, now we can use equation \eqref{formula_lyapunov_vert}. It will be enough to evaluate the quotient 
\begin{equation}
\frac{1}{t_k}\log\left(\frac{\Vert\mathcal{Q}'(t_k)G'(t_k-t_{k-1})...\mathcal{Q}'(t_1)G'(t_1)\delta\bm\eta(0)\Vert}{\Vert\delta\bm\eta(0)\Vert}\right),
\label{aprox_lya}
\end{equation}
for a large enough $k\in\mathbb{N}$.

\section{Application to Nakano-Saito oscillator}
\label{seccion_acoples}
Now we consider \emph{hybrid} networks of oscillators which have the form $\dot{\bm x} = A\bm x$. These networks generalize the \emph{dummy slave} exposed in the first section. We study the case of nodes coupled in different ways given by a matrix $E$. 

First we consider the classical coupling used as canonical test for evaluations of MSF. After we consider another coupling which if of interest in the design of electronic circuits.

In addition, we will take advantage of the stability of the simulations in order to investigate in detail different types of couplings, and we will see how those that privilege the voltage variable are fundamental in the generation of stability regimes.

In the simulations make with formula \eqref{aprox_lya}, we take $k=50$ for obtaining the MSF map showed in the figures.
\subsection{The canonical case}
We study the \emph{hybrid} network of oscillators only coupled in the first variable by a laplacian double-balanced matrix $L$ . Consider
\begin{equation}
E= E_{11} =
\begin{pmatrix}
1 & 0\\
0 & 0
\end{pmatrix},\quad
F =
\begin{pmatrix}
1\\
0
\end{pmatrix},\quad
I =
\begin{pmatrix}
v_R\\
\rho (1-v_R)
\end{pmatrix}.
\end{equation}
Considering a network like \eqref{red_hibrida} with the parameters mentioned above, we obtain an hybrid network whose synchronization solution $\bm s(t)$ is a Nakano-Saito oscillator given by \eqref{sistema_nakano}. 
\begin{figure}
\includegraphics[scale = 0.4]{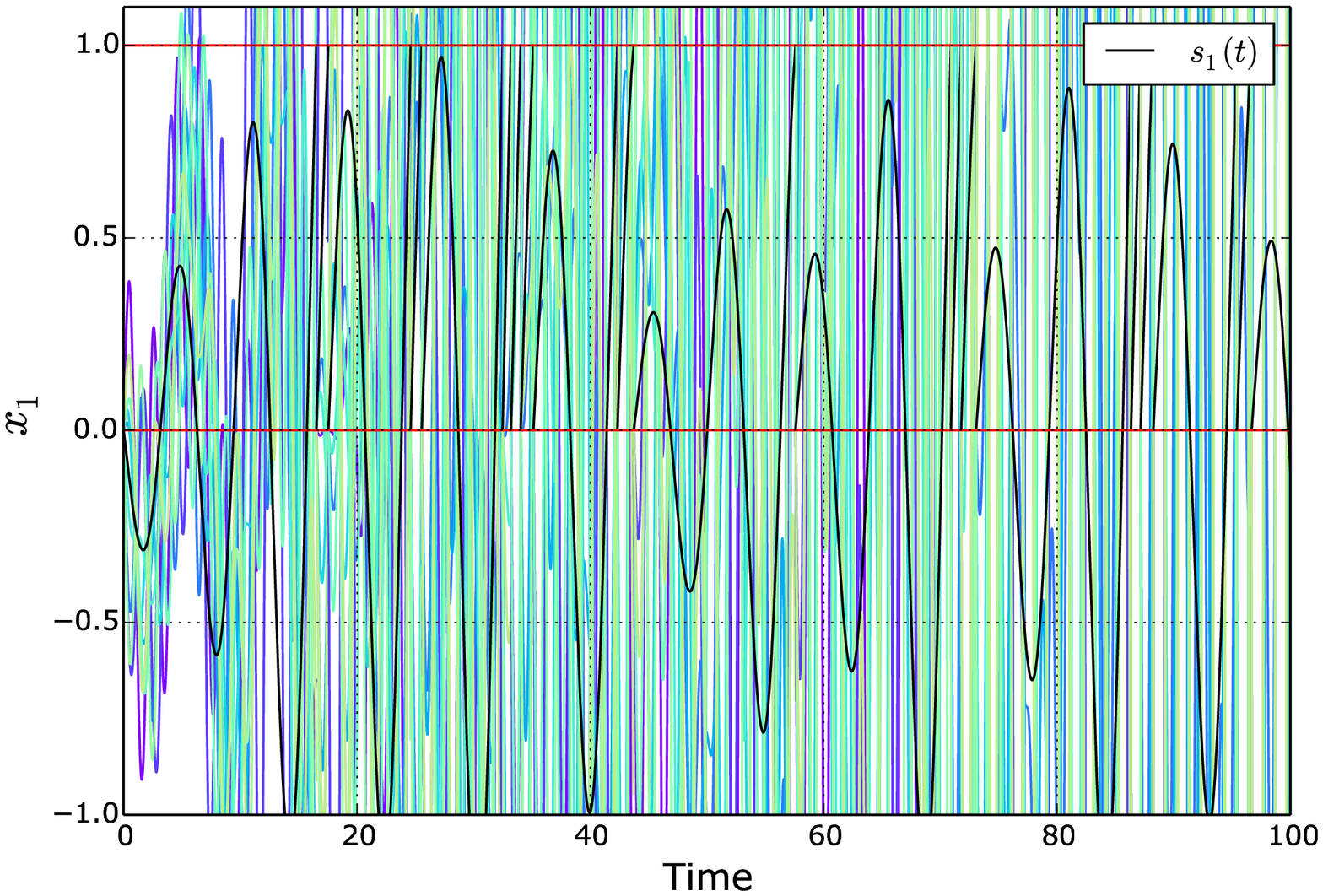}
\includegraphics[scale = 0.4]{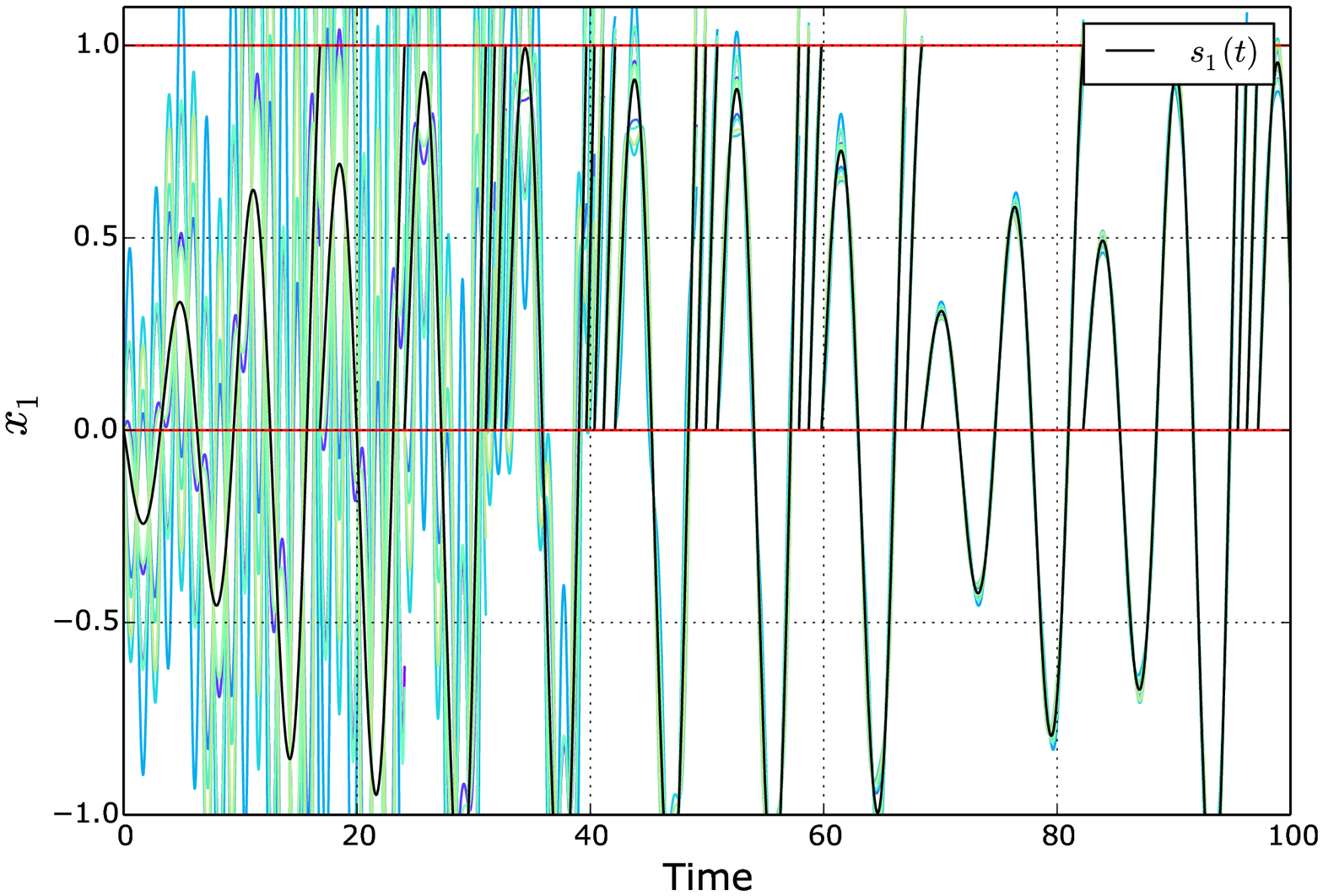}
\caption{Synchronous and non-synchronous behavior of the variable $\bm x_1$, of networks with a structure of complete non-directed graph (nine nodes), and random values of couplings. In black, the synchrony solution is plotted (parameters:$\rho = 0.1$, $v_R=0$).}
\label{comportamientos}
\end{figure}
As we can see in Figure \ref{comportamientos}, synchronous and non-synchronous behaviors are possible in these networks.

Applying the formalism of previous section, we can evaluate the MSF of the hybrid network \eqref{red_hibrida} for this oscillator.
\begin{figure}
\includegraphics[scale = 0.55]{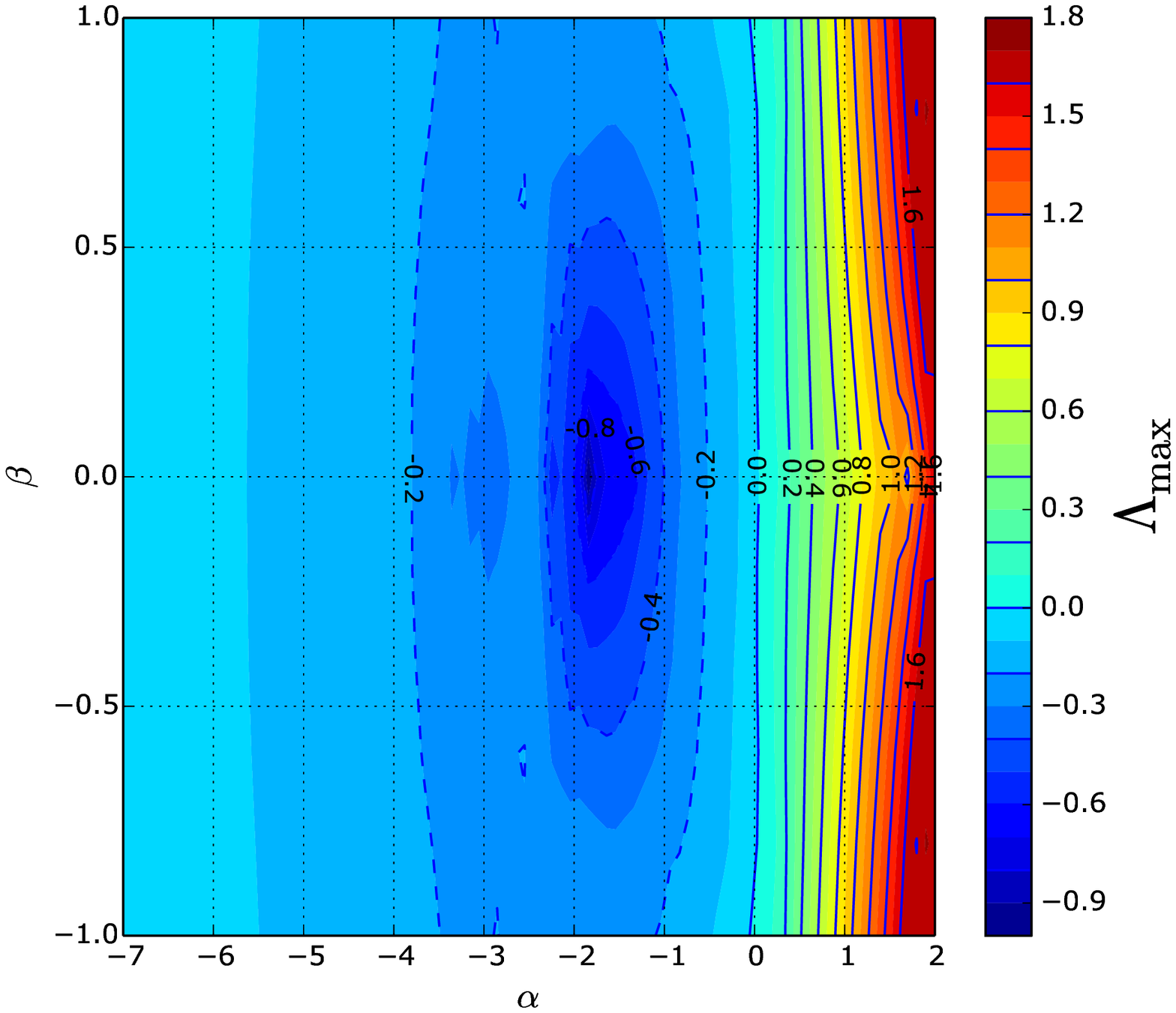}
\caption{MSF of the hybrid network with Nakano-Saito oscillator as synchronous state coupled by the matrix $E_{11}$ (parameters:$\rho = 0.1$, $v_R=0$).}
\label{MSF1_nakano}
\end{figure}
We can see in Figure \ref{MSF1_nakano} that the stability of the synchronous state is guaranteed for eigenvalues $\lambda$ with negative real part sufficiently far from zero. In addition, as is common in these maps, we can observe a well region (the dark blue region) where the synchrony is more robust.
\subsection{Coupling by the current variable}
Often systems like \eqref{circuito} are coupled by the current variables:
\begin{equation}
\mathcal{L}\frac{di_k}{dt}=-v_k+\mathcal{R}i_k+\sum_j w_{kj}(v_k-v_j).
\end{equation}
If we use the aforementioned dimensionless variables, we obtain an hybrid network like \eqref{red_hibrida} coupled by the matrix
\begin{equation}
E= E_{21} =
\begin{pmatrix}
0 & 0\\
1 & 0\\
\end{pmatrix}.
\end{equation}
This is one of the most interesting couplings, as we will see in the next section.
\begin{figure}
\includegraphics[scale = 0.55]{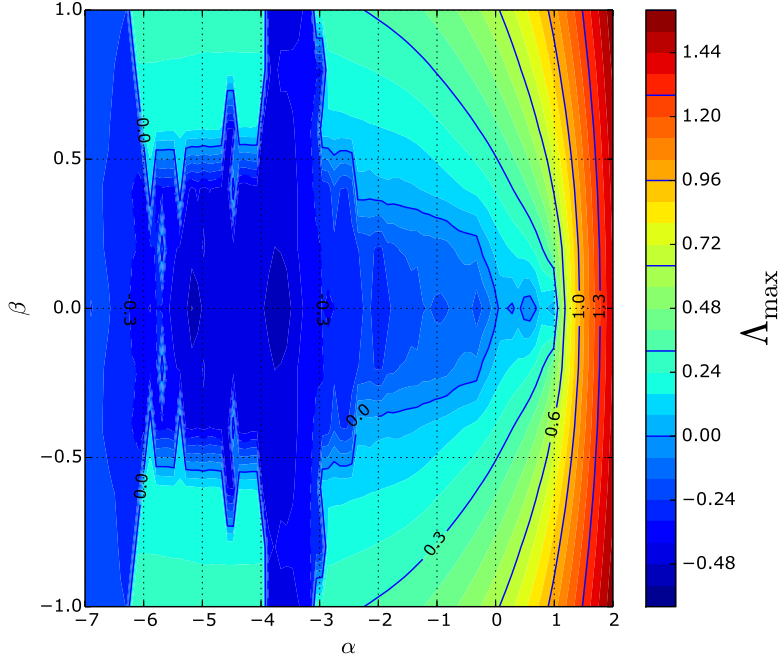}
\caption{MSF of the hybrid network with Nakano-Saito oscillator as synchronous state coupled by the matrix $E_{21}$ (parameters:$\rho = 0.1$, $v_R=0$).}
\label{MSF2_nakano}
\end{figure}

In Figure \ref{MSF2_nakano} we can observe a more complicated region of synchrony than in the previous example, with zones of the left half-plane where synchrony is not possible. Then we can conclude that the synchrony is more sensitive to the topology of the network for this type of couplings (since new regions of non-synchronization appear). Although again we see a well region (the dark blue region) where the synchrony is more robust.

\begin{figure}
\includegraphics[scale = 0.45]{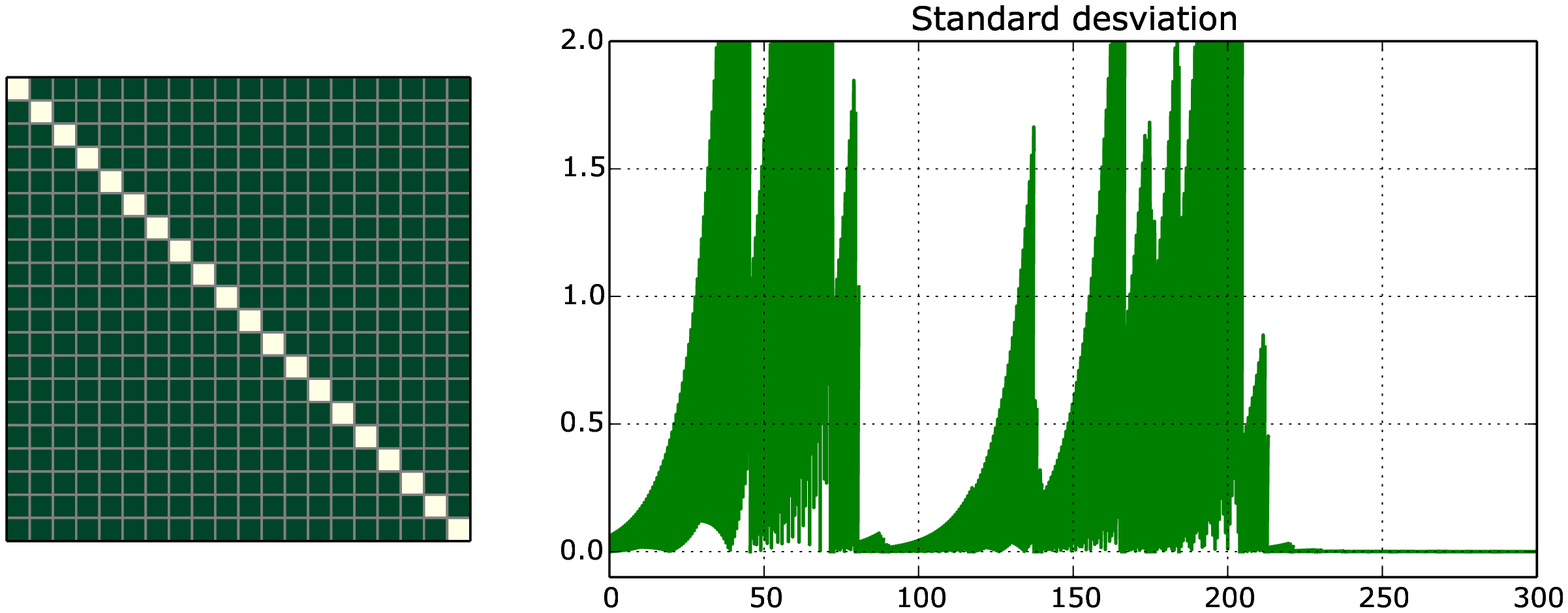}
\includegraphics[scale = 0.45]{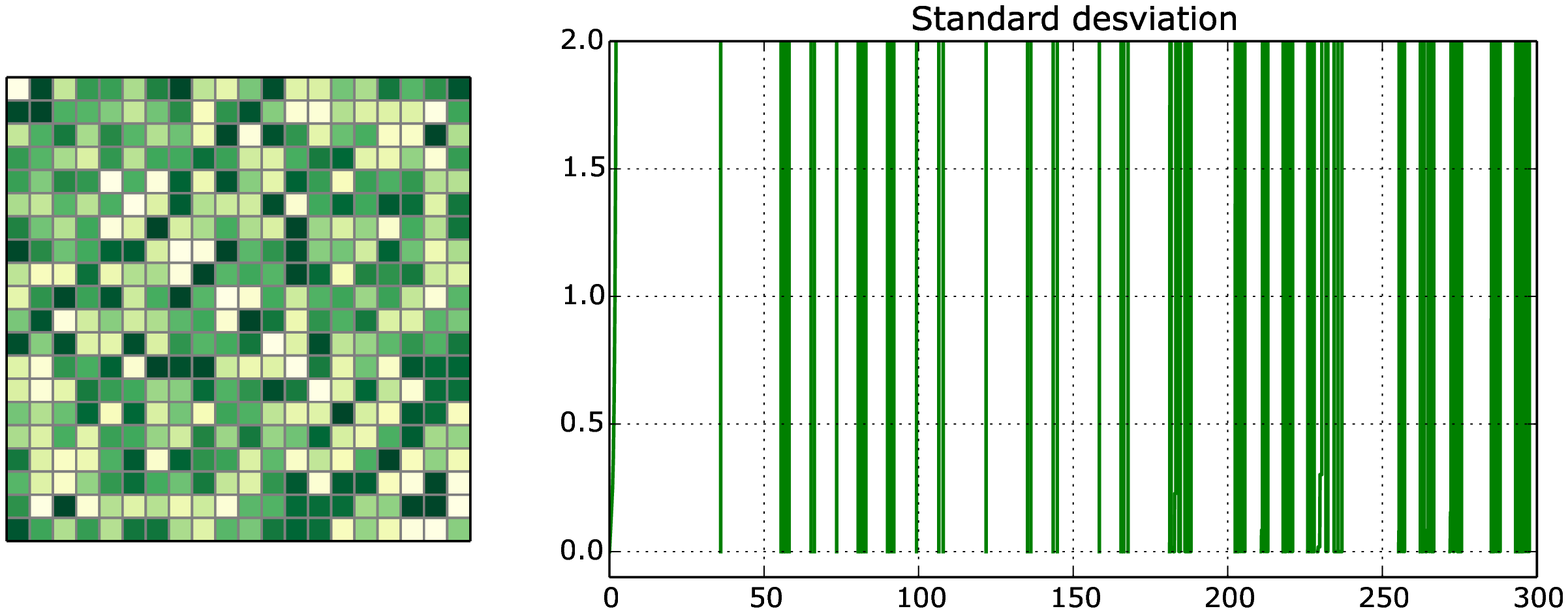}
\includegraphics[scale = 0.45]{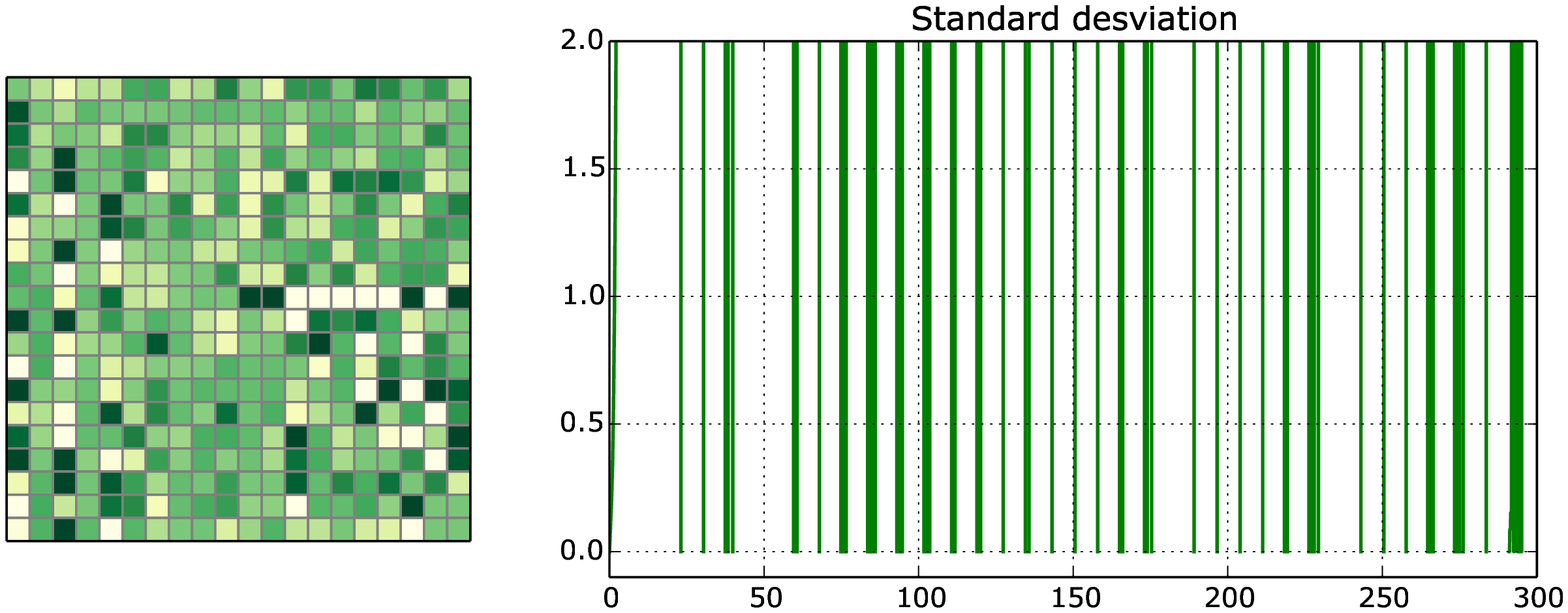}
\caption{Temporal evolution of the standard deviation of the first variable for different network structures, with connection $E_{21}$ (parameters:$\rho = 0.1$, $v_R=0$). The value of the standard deviation is close to zero near the synchronization state. The top image corresponds to the complete graph, and the $L$ matrix is its adjacency matrix, for the second image a symmetric matrix with random entries was used (the entries are in the interval $(-1,1)$ and the values of the diagonal entries are equal to the sum of the remaining entries of the row). The abrupt change to the synchrony regime with the complete graph is striking. Finally, the bottom image corresponds to a matrix balanced by both rows and columns, with random entries in $(-1,1)$ and null diagonal. In the pictorial representation of the matrices, the dark green shades correspond to values close to 1 and the light green shades to -1.}
\label{evoluciones}
\end{figure}

\begin{figure}
\includegraphics[scale = 0.55]{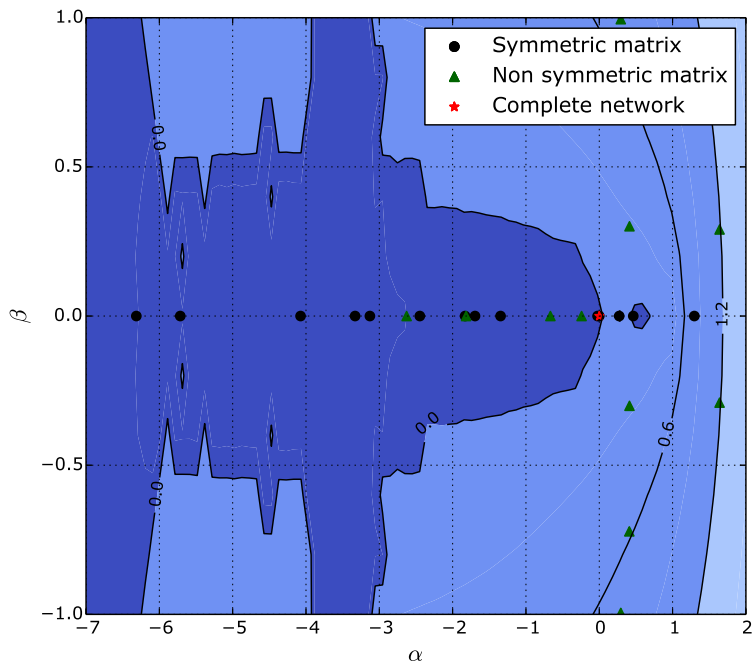}
\caption{The distribution of their eigenvalues is plotted in the figure, and it can be seen how the inside/outside position of the respective eigenvalues coincide with the synchrony simulations shown in the previous figures. The dark blue areas correspond to the synchrony regions of the network.}
\label{MSF_autovalores}
\end{figure}
Numerical simulations of hybrid networks with $E_{21}$ coupling are shown in Figure \ref{evoluciones}. Temporal standard deviation of the first variables of nodes has been taken as synchrony indicator, thus the case $\sigma = 0$ corresponds to the synchrony state. The network is simulated for three different configurations (case of complete network, symmetric random network and other random directed network). Figure \ref{MSF_autovalores} shows the distribution of the eigenvalues of the simulated network configurations together with the MSF. It can be appreciated the coincidence between the predicted behavior of the network and the simulated one.
\subsection{Other couplings} 
We will concentrate only on networks with non-directed connections, to study the impact of the $ E $ matrix that selects the coupling mode. In this case, the  matrix $L$ is symmetric, and all its eigenvalues are real. Thus, the evaluation of the MSF is simplified to case $\lambda\in\mathbb{R}$.

We will consider different couplings: the family of the $E_{ij}$ matrices (which possess the $ij$-entry equal to 1 and zero the others entries), the identity coupling $E=\mathbb{I}_2$, and the coupling through the following matrix
\begin{equation}
G =
\begin{pmatrix}
0 & 1\\
1 & 0\\
\end{pmatrix}.
\end{equation}

\begin{figure}
\includegraphics[scale = 0.5]{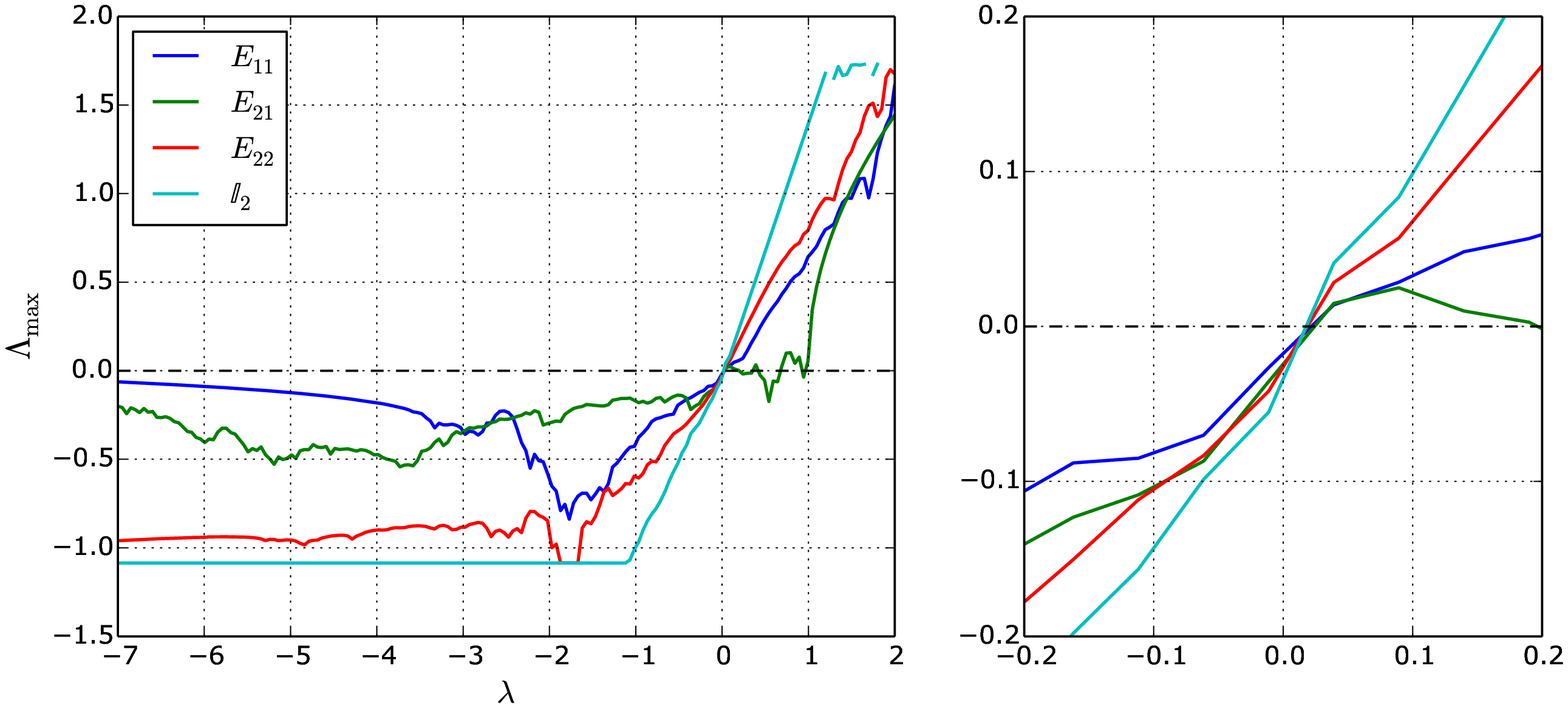}
\caption{Sectional cut of the MSF ($\Im(\lambda)=0$) for the couplings $E_{11},E_{21},E_{22}$ and $\mathbb{I}_2$. The right graph shows how the stability region begins just before the zero value. We can also identify in each curve some well region, as is usual in these diagrams. Finally we see that in the case of the coupling $E_{21}$ there are others stability zone to the right of $\lambda = 0$.}
\label{couplings_1}
\end{figure}

The simulations of the MSF corresponding to the couplings $E_{11},E_{21},E_{22}$ and $\mathbb{I}_2$ were performed using the formula \eqref{aprox_lya} with $K=50$. In all these cases, near zero there is a change in stability the synchronization solution from right to left along the real axis. The horizontal part of the curve corresponding to the coupling $\mathbb{I}_2$ is due that in the numerical implementation the value of $\Vert\delta\bm\eta_k\Vert$ was lower than the allowed tolerance $\epsilon = 10^{-100}$. So, that value was taken as an estimate.
In this way, the constant part of the curve does not match the correct value of $\Lambda_{\max}$ (in fact its value is much smaller) but it does guarantee its negative sign. The plot of the MSF corresponding to that couplings can be seen in Figure \ref{couplings_1}. Finally, we compare the curves and obtain that the coupling $E_{21}$, unlike the others, presents disjoint intervals of stability, which allows to infer a richer behavior of the networks that use it.

\begin{figure}
\includegraphics[scale = 0.5]{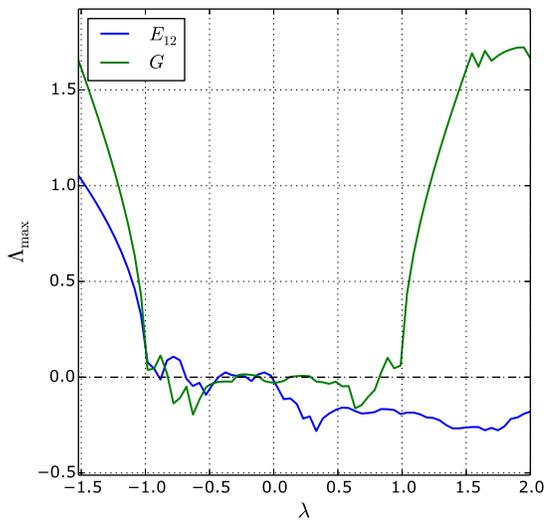}
\caption{Sectional cut of the MSF ($\Im(\lambda)=0$) for the couplings $E_{12}$ and $G$. We can see how in the coupling $E_{12}$, the appearance of the stability region occurs as the value $\lambda$ increases, while in the case of $ G $ there is a "gap" of stability between the values $(-1,1)$. In turn, we observe that stability changes are much more abrupt than in the other couplings considered.}
\label{couplings_2}
\end{figure}

On the other hand, the MSF of the couplings $E_{12}$ and $G$ could not be estimated in the same range as the previous ones due to the numerical instability they presented. However, the calculation could be performed in a smaller neighborhood of $\lambda = 0$ with the same precision as the previous ones. The direction of stability changes in these two couplings unlike the previous couplings. In Figure \ref{couplings_2} we can see how the instability region ends to the right of the value $\lambda = 0 $ and the stability region continues as the parameter grows.

We conjecture that such a change of direction is due to the influence of the variable $\bm x_2$ on $\bm x_1$ because the growth of the variable $\bm x_2$, unlike $\bm x_1$, is not restricted by any threshold and the reset does not change its magnitude. Thus, their influence on $\bm x_1$ tends to increase the frequency with which resets occur, destabilizing the network (we can see that the slope in the changes of the stable regions is much greater than in the previous couplings).

\section*{Conclusions}
In this work the appearance of synchronous behaviors in a particular type of pulsed coupled networks with global feedback was studied. Like classic synchronization works, our network consists of coupled identical nodes. Each node of network represents a Nakano type circuit, and they are linearly coupled. However, the action of a common reset to the entire network (given by the average of the voltage variables) produces a discontinuous dynamics of integrate-and-fire type.

Due to certain characteristics of this network, several of the usual formalism (mainly the MSF and the saltation matrix) could be generalized and expressed through closed forms.

In this way it was possible to investigate how a global reset rule generates synchrony states and how the reset regimes can be canceled under certain hypotheses about the topology of network. Also different coupling modes can be compared. On this last point, it should be noted that after comparisons of different coupling modes, those involving the voltage variables acting on the current ones stand out. This type of coupling exhibits a more robust synchrony and the appearance of different stability gaps. This variety of behaviors makes this coupling the candidate to develop control mechanisms for this type of networks.

However, it is still pending to explain the nature of these behaviors, as well as to expand these formalism to more complex and higher dimensional systems than the Nakano circuit. On the other hand, although taking the average of the variables as restart, allowed us to study in detail the synchrony, in future works the rule of restart must be replaced by more realistic ones. Some of the possible couplings are random choices of nodes, connections with delay \cite{kinzel09} or linear functions such like those used in \cite{burbano16}.

Finally, all these studies will also involve sophisticated numerical tools. Although the MSF is a tool that provides invaluable information on network dynamics, in future studies we will incorporate new approaches such as phase curve response or mean-field quantities, which allow us to account for different behaviors such as the transition between synchrony states and asynchrony \cite{pecora98}, anticipated synchrony \cite{voss2000anticipating}, or traveling waves through the network connections \cite{izhikevich08}.
\section*{Acknowledgments}
The work is supported by the Universidad Nacional del Sur (Grant no. PGI 24/L096). The first author is also supported by CONICET.

\end{document}